\documentclass[]{elsarticle}
\usepackage[a4paper, total={6in, 8in}, margin=0.9in]{geometry}
\usepackage{multicol}
\usepackage{amssymb}
\usepackage{graphicx}
\usepackage{booktabs}
\usepackage{multirow}
 
\usepackage{amsmath}
\usepackage{caption}
\usepackage{subcaption}
\usepackage{algorithm}
\usepackage{algpseudocode}

\usepackage{amsmath,amssymb,amsfonts}
\usepackage{algorithm}
\usepackage{algpseudocode}
\usepackage{graphicx}
\usepackage{textcomp}
\usepackage{multirow}
\usepackage{longtable}
\usepackage{rotating} 
\usepackage{longtable}
\usepackage{booktabs}
\usepackage{ltxtable} 
\usepackage{supertabular,booktabs}
\usepackage[export]{adjustbox}
\usepackage{tablefootnote}
\usepackage{color}
\usepackage{xcolor}
\usepackage{url}

\usepackage{changes}

\begin{document}

\begin{frontmatter}



\title{Class-Aware Adaptive Differential Privacy in Deep Learning for Sensor-Based Fall Detection}


\author[ss]{Joydeb Kumar Sana}
\address[ss]{Institute of Information and Communication Technology, Bangladesh University of Engineering and Technology, Dhaka, Bangladesh (e-mail: joysana@gmail.com, joydebsana@iict.buet.ac.bd)}

\cortext[cor]{Joydeb Kumar Sana}
\ead{joysana@gmail.com, joydebsana@iict.buet.ac.bd}





\author{  }

\begin{abstract}
Fall detection is a critical task in healthcare, particularly for elderly people. Timely fall detection and treatment can prevent severe injuries. Sensor-based activity data can be used to detect fall. However, this data are highly sensitive and raises significant privacy concerns. Existing privacy approaches apply uniform noise across all training samples, which affects the prediction performance. To address this limitation, we propose a Class-Aware Adaptive Differential Privacy (CA-ADP) framework integrated with a hybrid 3D Convolutional Neural Network and Bidirectional Long Short-Term Memory (3D CNN-BiLSTM) architecture. The CA-ADP mechanism dynamically adjusts the magnitude of noise added to gradients based on the class composition of each mini-batch. This process ensures privacy while mitigates performance degradation. We formally analyze the $(\epsilon,\delta)$-Differential Privacy guarantee and provide a privacy-utility trade-off analysis. The proposed method is evaluated on three public benchmark datasets, namely SisFall, UP-Fall, and MobiAct. The experimental results show that the proposed privacy model achieves improvements of 3.3\%, 8.5\%, and 7.5\% over the conventional privacy-based model in terms of F-score for the SisFall, UP-Fall, and MobiAct datasets, respectively. Comparisons with prior studies show that the CA-AD based framework achieves competitive performance and provides formal privacy guarantees, which are largely overlooked in existing studies. Wilcoxon signed-rank tests confirm that the proposed mechanism consistently outperforms conventional differential privacy. Those results establish the proposed CA-ADP framework as an effective approach to privacy-preserving fall detection in real-world healthcare settings.
\end{abstract}


\end{frontmatter}

\section{Introduction}

Falls are a major cause of injury and hospitalization among elderly people around the world. According to the World Health Organization, falls are a major public health concern for aging people. Timely and reliable fall detection system can significantly reduce response time, mitigate complications, and improve patient outcomes. Advanced wearable and smartphone-based sensing technologies can be used for continuous monitoring of human activities through accelerometers and gyroscopes \cite{SisFall_2017, Maruf_2025}. These sensor-based systems give several advantages over vision-based methods, including lower cost, portability, and reduced environmental dependency. Deep learning models such as Convolutional Neural Networks (CNNs) and Recurrent Neural Networks (RNNs) have demonstrated their effectiveness in human activity recognition (HAR) tasks \cite{Pathan_2025}. Recent studies ~\cite{Fan_2024, khatun2025multiagent} show that hybrid architectures, such as CNNs integrated with Bidirectional Long Short-Term Memory (BiLSTM) networks, can achieve promising results for motion classification problems. These hybrid models usually combine spatial features and temporal modeling. Despite their effectiveness, deep learning-based fall detection systems raise significant privacy concerns~\cite{abadi2016deep}. Trained models may suffer from privacy attacks and adversaries may be able to collect sensitive information from the trained models ~\cite{JDK_2025_PPCCP}. Sensor data collected from wearable devices may contain sensitive behavioral and health-related information. In healthcare, protecting user privacy is not optional but essential. Researcher uses Differential Privacy (DP) \cite{DworkCynthia2006} to protect user data. Differential Privacy (DP) is a mathematical framework that offers formal guarantees against information leakage. Differentially Private Stochastic Gradient Descent (DP-SGD)\cite{abadi2016deep} is a well-known DP mechanism. DP-SGD uses gradient clipping and calibrated noise during training to protect individual records. However, conventional DP-SGD (CONV-DP) approaches often apply uniform noise across all training samples and classes.  This can be a problem with datasets that are imbalanced, which are common in fall detection scenarios. Uniform privacy mechanisms may disproportionately affect minority classes and degrade model performance.  

To address these limitations, this paper proposes an Adaptive Differential Privacy framework for sensor-based fall detection using a hybrid 3D CNN–BiLSTM architecture. The proposed method introduces adaptive noise scheduling during training based on the class distribution. It allows the process to dynamically adjust noise intensity across mini-batches. This strategy aims to preserve privacy guarantees while maintaining or improving model performance. The main contributions of this work are summarized as follows:

\begin{itemize}
\item We propose a novel Class-Aware Adaptive Differential Privacy (CA-ADP) mechanism that dynamically adjusts noise injection based on mini-batch class composition to improve privacy–utility trade-offs.

\item We develop a hybrid 3D CNN–BiLSTM architecture for sensor-based fall detection, capable of capturing both spatiotemporal features and long-term temporal dependencies.

\item We integrate the proposed CA-ADP mechanism into the 3D CNN–BiLSTM framework to enable privacy-preserving training without significantly degrading model performance.

\item We conduct a comprehensive empirical analysis of the privacy–utility trade-off under adaptive differential privacy.

\item We evaluate the proposed approach on three publicly available benchmark datasets using multiple evaluation metrics to demonstrate its effectiveness and robustness.

\item To the best of our knowledge, this is the first study to apply Class-Aware Adaptive Differential Privacy mechanism to fall detection.
\end{itemize}

\section{Background Study} \label{sec:literature_review} 

Fall detection has been broadly studied within the domain of Human Activity Recognition (HAR). Early fall detection approaches mainly relied on threshold-based methods~\cite{Bourke_2008}. These techniques used accelerometer signals, where sudden spikes in acceleration were interpreted as fall events. These methods are computationally simple. However, they often produce a high number of false alarms and struggle to adapt to more complex movements. In the Machine Learning (ML) era, several ML classifiers, such as Support Vector Machines (SVMs), k-Nearest Neighbors (kNN) and Random Foress (RF) have been  applied to wearable sensor data~\cite{Ishak_2021}. These approaches improved classification performance. However, they required handcrafted feature engineering which is time consuming and very subjective. Currently, researchers are using deep learning techniques instead of traditional approaches because of their ability to automatically learn hierarchical features from raw data. In~\cite{Fakhrulddin_20217},  Convolutional Neural Networks (CNNs) have been used to detect human activity on sensor data. Gunale et al. in \cite{Gunale2023CNN} also show the advantage of using CNN for fall detection. Long Short-Term Memory (LSTM) networks can also enhance performance. They have the potential to capture temporal dependencies in sequential data~\cite{Uddin_2021, ordonez2023deep}. Recurrent neural network (RNN) with LSTM architecture shows its effectiveness in detecting fall events \cite{Hasan2019RNN}. Hybrid CNN-LSTM architectures show promising prediction performance. They combine spatial feature extraction with sequential modeling, which makes them suitable for fall detection tasks~\cite{HU_2025}. Most recent studies explore the potential of 3D CNNs and multimodal sensor fusion techniques. They achieve advantages by capturing spatial and temporal correlations ~\cite{Alanazi2022, Almukadi2024}. These approaches have significantly improved fall recognition accuracy and robustness. Despite these developments, most existing fall detection methods primarily focus on improving performance while overlooking the privacy risks associated with sensor-based human activity monitoring. Differential privacy (DP) can provide a practical solution to overcome this challenge by protecting sensitive information in the training data.

 Differential Privacy (DP) is a protective way of data privacy. This method was formally introduced by Dwork ~\cite{DworkCynthia2006}. In practice, DP protects privacy by injecting noise. However, noise injection can affect model prediction performance. To balance between privacy and performance, different DP mechanism follows different strategy. In deep learning, DP is commonly implemented through Differentially Private Stochastic Gradient Descent (DP-SGD). It clipes per-sample gradients and incorporates uniform Gaussian noise to preserve privacy~\cite{abadi2016deep}. This noisy gradients are then used for parameter updates during training. DP-SGD has been used in various deep learning applications, including image classification and natural language processing~\cite{Wei_2025}. Though DP provides strong theoretical guarantees, it often introduces a trade-off between privacy and model performance. Higher noise levels provide better privacy but may also degrade prediction accuracy~\cite{PAN_2024}. To address this limitation, Several recent works have investigated adaptive privacy mechanisms, such as dynamic noise scaling and privacy budget scheduling~\cite{Andrew_2021, Zhang_2026}. These approaches aim to reduce unnecessary noise injection during the training process. Adaptive DP strategies have demonstrated improved utility in various domains. However, their application to sensor-based fall detection remains largely unexplored.




Healthcare monitoring systems collect personal data to support patient care, much of which contains sensitive medical details. Leveraging this sensitive data for training triggers privacy concerns. Beyond the ethical implications, this is a mandatory obligation~\cite{Guerra_Manzanares_2023, Raju_2024}. Trained model can leak training information which allows adversaries to extract sensitive personal records. Although encryption effectively protect data during transmission and storage, it can not prevent information leakage during model training and deployment stages~\cite{abadi2016deep}. Differential Privacy (DP) addresses this limitation by limiting the impact of any single record on the model. Even with its strong theoretical backing, the applications of DP in fall detection research are limited compared to performance-focused approaches. A further limitation is that conventional DP applies uniform noise across all training records, which disproportionately affects the minority class and ultimately degrades prediction performance~\cite{pichapati2019adaclip, choi_2026}. This gap motivates the need for class-aware adaptive privacy mechanisms tailored to imbalanced healthcare datasets.

\subsection{Research Gap}

We found several critical gaps that need to be addressed:
\begin{itemize}
 \item Most existing fall detection approaches focus on improving classification accuracy, with limited consideration of data privacy.
 \item Most differential privacy methods still rely on uniform noise injection, which does not consider the strong class imbalance in fall detection datasets.  As a result, the performance on minority (fall) cases can be disproportionately affected.
 \item There is a lack of integrated frameworks that jointly address spatiotemporal feature learning and privacy preservation while maintaining a favorable privacy–utility trade-off.
\end{itemize}
To address these limitations, this paper proposes a class-aware adaptive differential privacy framework integrated with a hybrid 3D CNN–BiLSTM architecture. The aim of this research is to improve fall detection performance while ensuring data privacy.

\section{Materials and Methods}\label{sec:methodology}
This section describes the detail overview of the datasets, data preprocessing, class-aware differential privacy techniques employed in this study. It also presents the proposed deep learning framework, evaluation measures, statistical test, and other relevant topics.

\subsection{Datasets} \label{sec:Datasets} 
 To comprehensively evaluate the proposed Class-Aware Adaptive Differential Privacy (CA-ADP) framework, experiments were conducted on three widely used public benchmark datasets for sensor-based fall detection: SisFall~\cite{SisFall_2017}, UP-Fall~\cite{UP_Fall_2019}, and MobiAct~\cite{MobiAct}. These datasets vary in subject demographics, sensor placement, sampling frequency, and activity diversity, providing a heterogeneous and realistic evaluation setting for privacy-preserving deep learning models.
  

\subsubsection{SisFall Dataset}
The SisFall dataset~\cite{SisFall_2017} is a well-known benchmark for fall detection using accelerometer and gyroscope data. This dataset is designed to study falls among older adults. It contains data collected from 38 subjects, including both young and elderly participants. The Data were collected using wearable inertial measurement units (IMUs) placed at the waist. It includes 15 types of simulated fall activities and 19 types of Activities of Daily Living (ADLs). Sensor signals were recorded at a high sampling frequency of 200 Hz. Each sample contains six-dimensional data from tri-axial accelerometers and gyroscope. It is a class imbalance dataset. The total number of non-fall activities (2707) is larger than fall events (1798). Despite this imbalance, SisFall provides realistic fall scenarios and diverse daily activities. It is suitable for evaluating model robustness and generalization ability of fall detection models in real-world settings.

\subsubsection{UP-Fall Dataset}
Muñoz et al. \cite{UP_Fall_2019} developed UP-Fall dataset to support fall detection research. It includes recordings from 17 subjects performing both fall events and Activities of Daily Living (ADLs) under controlled conditions. As our focus is on privacy-preserving, sensor-based modeling, we consider only the wearable sensor modality and exclude the RGB and depth data provided in the dataset. The dataset has six daily living activities and five types of fall activities. Accelerometer and gyroscope sensors record three-dimensional inertial data across dynamic movements. While the dataset covers fewer activity categories and participants than SisFall, its standardized execution protocol offers a well-controlled experimental foundation.

\subsubsection{MobiAct Dataset}
The MobiAct dataset \cite{MobiAct} \cite{MobiAct_2} is a smartphone-based activity recognition dataset. It aims to mimic real-world fall detection through mobile devices. It has recordings of 57 people doing both daily activities and fall events. MobiAct uses the embedded sensors of smartphones, generally carried in the front trouser pocket. This setup makes the data more realistic, but variations in phone orientation and posture introduces more fluctuation. The dataset includes four types of falls and nine daily activities. It collects data from the smartphone’s accelerometer, gyroscope, and orientation sensors. All signals are sampled at 20 Hz. Similar to the SisFall and UP-Fall datasets, this dataset is also imbalanced, where fall events represent a minority class relative to daily activities.

\section{Data Preprocessing} \label{subsec:datapreparation}
Effective data preprocessing is a critical step in developing a robust system. In this fall detection research, we used heterogeneous public datasets such as the SisFall, UP-Fall and MobiAct. Though these datasets differ in several characteristics, we designed a standardized preprocessing pipeline to ensure consistency. Our goal is dataset specific optimization, not cross dataset fairness. Therefore, the window size was chosen based on the sampling frequency and signal properties. The datasets were split into 80\% training and 20\% testing sets using stratified sampling. Several Python libraries were used to facilitate the data preprocessing pipeline.

\subsection{SisFall Dataset}
The SisFall dataset is organized into subject-specific directories. Each activity is saved as a separate text file. The preprocessing steps were implemented as follows:

\begin{itemize}
    \item  Columns containing missing values were removed.
    
    \item If a time column was present (i.e., total 7 columns), it was removed so that only the relevant inertial sensor data remained.
    
    \item To reduce noise, a 4\textsuperscript{th}-order Butterworth low-pass filter was applied to each signal channel.
    \item With a sampling rate of 200 Hz and a cutoff frequency of 20 Hz, zero-phase filtering was employed to remove high-frequency noise.

    \item The filtered signals were normalized for each channel. 
    
    \item The continuous signals were then segmented using a sliding window of 200 samples with a 50\% overlap (step size of 100 samples).
    
    \item Binary labels were assigned based on file naming: files starting with ``F'' were labeled as fall (1), while all others were labeled as non-fall (0).

\end{itemize}

\subsection{UP-Fall Dataset}

The UP-Fall Dataset was loaded from the \texttt{CompleteDataSet.csv} file. The following preprocessing steps were applied:

\begin{itemize}
    \item Feature columns (1 - 18) and the label column were separated.
    
    \item All values were converted to numeric format. Rows containing invalid or missing values were removed.
    
    \item Forward-fill imputation was applied where necessary to maintain signal continuity.
    
    \item Feature normalization was performed using StandardScaler.
        
    \item Multi-class labels were converted into binary classes: labels 1--5 were mapped to fall (1), and the remaining labels to non-fall (0).

    \item The normalized features and binary labels were segmented using a sliding window approach with a window size of 50 samples and a step size of 25 samples, corresponding to a 50\% overlap between consecutive windows. Each window was assigned the statistical mode of the labels within it.

\end{itemize}

\subsection{MobiAct Dataset}

The MobiAct Dataset (v2.0) was loaded from the Annotated Data directory, where each folder corresponds to a specific activity type.
\begin{itemize}
    \item Fall activities were identified from the folders \texttt{FOL, FKL, FKR, SDL, SDR, BSC}. Files in these folders were labeled as fall (1), while all others were labeled as non-fall (0).
    
    \item From each CSV file, sensor columns 1--6 were extracted to construct multivariate time-series data.
    
    \item Sliding window segmentation was performed with a window size of 128 samples and a step size of 64 (50\% overlap).
    
    \item Features were normalized using \texttt{StandardScaler} after reshaping the data.

\end{itemize}

This preprocessing pipeline transforms the different datasets into a consistent format. Table~\ref{tab:preprocessing_summary} provides a overview of the preprocessing configuration applied to each dataset.

\begin{table}[h]
\centering
\caption{Summary of Preprocessing Configuration Across Datasets}
\label{tab:preprocessing_summary}
\begin{tabular}{|l|c|c|c|}
\hline
\textbf{Configuration} & \textbf{SisFall} & \textbf{UP-Fall} & \textbf{MobiAct} \\
\hline
Data Source & Subject-wise .txt files & CompleteDataSet.csv & Activity-wise folders (.csv) \\
\hline
Sensor Type & Wearable IMU & IMU (selected) & Smartphone IMU \\
\hline
Sampling Rate & 200 Hz & Not explicitly used & Not explicitly used \\
\hline
Filtering & 4th-order Butterworth LPF (20 Hz) & None & None \\
\hline
Normalization & StandardScaler & StandardScaler & StandardScaler \\
\hline
Window Size & 200 samples & 50 samples & 128 samples \\
\hline
Step Size & 100 samples & 25 samples & 64 samples \\
\hline
Overlap & 50\% & 50\% & 50\% \\
\hline
Label Strategy & File prefix (F = Fall) & Labels 1--5 = Fall & Folder-based mapping \\
\hline
Binary Mapping & Fall=1, ADL=0 & Fall=1, ADL=0 & Fall=1, ADL=0 \\
\hline
Train-Test Split & 80:20 (Stratified) & 80:20 (Stratified) & 80:20 (Stratified) \\
\hline
\end{tabular}
\end{table}

\subsection{Differential Privacy}
Differential Privacy (DP) \cite{DworkCynthia2006} is a privacy protective mechanism of personal data. It is a set of conditions that must be satisfied to ensure the privacy of the sensitive data. In general, DP injects random noise in the training process or query that does not influence the outcome of any useful analysis. It makes an obstacle to the adversary. The core principle of DP is that an adversary should not be able to reliably distinguish between two datasets that differ by only one record. In other words, the presence or absence of a specific individual's data should not significantly affect the output of the analysis. DP achieves a balance between privacy and utility through a privacy budget parameter, commonly denoted as $\epsilon$. A smaller value of $\epsilon$ provides stronger privacy guarantees but may reduce model accuracy due to increased noise injection.

In practical scenarios such as fall detection, activity recognition, and medical diagnosis, datasets often contain highly sensitive personal information. Direct training of machine learning models on such data may lead to privacy leakage. The trained model may suffer from membership inference attacks, model inversion attacks, or unintended memorization of sensitive patterns. DP provides  formal guarantees that limit the risk of exposing individual-level information. The DP is first introduced by the Dwork \cite{DworkCynthia2006}, where he presented a mathematical definition that ensures data privacy. DP can be defined as follows.

A randomized mechanism $\mathcal{M}$ satisfies 
$(\epsilon,\delta)$-Differential Privacy if for any two neighboring datasets $\mathcal{D}$ and $\mathcal{D}'$ differing in at most one sample, and for any possible output set $S$:

\begin{equation}
P [\mathcal{M}(\mathcal{D}) \in S]
\leq
\exp(\epsilon) \cdot 
P [\mathcal{M}(\mathcal{D}') \in S]
+
\delta
\end{equation}

where, $S$ denotes any possible subset of outputs of the mechanism $\mathcal{M}$, and $P$ represents the probability distribution  induced by the randomized mechanism $\mathcal{M}$. The parameter  $\epsilon$, known as the privacy budget, controls the strength of the privacy guarantee: smaller values of $\epsilon$ provide stronger privacy protection, whereas larger values allow higher utility at  the cost of reduced privacy \cite{Lee2011, xie2018differentially}. The parameter $\delta$ represents a small probability of failure of the strict privacy guarantee. While $\epsilon$ controls the maximum multiplicative difference between the output distributions of two  neighboring datasets, $\delta$ allows for a small relaxation.  In other words, with probability at most $\delta$, the privacy loss may exceed the bound determined by $\epsilon$.

In this mechanism, the injected random noise must be sampled from a probability distribution which satisfies the DP requirements. The sampled noises are then added to the query of individual data records. The amount of noise required is directly proportional to the sensitivity of the query function. Therefore, the noise magnitude must be sufficiently large to obscure the impact of any individual record, ensuring that the presence or absence of a specific sample cannot be reliably assumed. Several mechanisms have been proposed to achieve Differential Privacy, including Laplace mechanism \cite{Geng_2016}, Gaussian mechanism \cite{Jinshuo2022}, Exponential mechanism \cite{Frank2007}, Private Aggregation of Teacher Ensembles (PATE) \cite{Jordon2019PATEGANGS}, etc. The Laplace mechanism is used for achieving pure DP, particularly in low-dimensional statistical queries. The Exponential mechanism is suitable for non-numeric outputs. PATE is designed for privacy-preserving deep learning through teacher–student knowledge transfer. Among these approaches, the Gaussian mechanism is widely used in modern deep learning frameworks due to its smooth noise distribution and flexible calibration \cite{Canonne_Kamath_Steinke_2022, xie2018differentially, Dwork2014}. Unlike the Laplace mechanism, which ensures pure DP, the Gaussian mechanism supports relaxed $(\epsilon,\delta)$-DP guarantees, making it more practical for large-scale optimization problems such as stochastic gradient descent. The continuous and symmetric nature of Gaussian noise also leads to more stable training behavior in neural network optimization. Therefore, in this research, we adopt a Gaussian mechanism-based Differential Privacy approach, where calibrated Gaussian noise is added during the training process to protect sensitive information while maintaining acceptable model performance.

The Gaussian mechanism achieves $(\epsilon,\delta)$-DP by adding zero-mean Gaussian noise to the output of a function computed on the dataset. Let $f(\mathcal{D})$ be a real-valued query function with $\ell_2$-sensitivity $\Delta_2 f$, defined as the maximum change in the function output when one sample in the dataset is modified. The Gaussian mechanism is defined as:

\begin{equation}
\mathcal{M}(\mathcal{D}) 
= 
f(\mathcal{D}) + \mathcal{N}(0, \sigma^2)
\end{equation}

where $\mathcal{N}(0, \sigma^2)$ denotes Gaussian noise with mean zero and standard deviation $\sigma$. To ensure $(\epsilon,\delta)$-DP, the noise scale is chosen as:

\begin{equation}
\sigma = \frac{\Delta_2 f \sqrt{2 \ln(1.25/\delta)}}{\epsilon}.
\end{equation}

Thus, the Gaussian mechanism ensures formal $(\epsilon,\delta)$-DP by calibrating the magnitude of the injected noise according to the function sensitivity, privacy budget $\epsilon$, and allowable failure probability 
$\delta$.

\subsection{Class-Aware Adaptive Differential Privacy}

In fall detection datasets, class imbalance is common, where fall events represent a minority class compared to non-fall activities. Standard DP-SGD applies uniform clipping and noise across all samples, which may disproportionately affect minority classes. To address this limitation, we introduce a Class-Aware Adaptive Differential Privacy (CA-ADP) mechanism that dynamically adjusts gradient clipping and noise injection based on class distribution, while maintaining a formal $(\epsilon, \delta)$-DP guarantee.







\subsubsection{Class Distribution}

Let $B = \{(x_i, y_i)\}_{i=1}^{|B|}$ be a mini-batch where $y_i \in \{0,1\}$ and $y_i = 1$ denotes a fall event. The mean batch label is computed as:

\begin{equation}
    \bar{y}_b \;=\; \frac{1}{|B|}\sum_{i \in B} y_i \;\in\; [0,1],
    \label{eq:batch_mean}
\end{equation}

\noindent which serves as a continuous proxy for the proportion of fall samples in the current batch. When $\bar{y}_b \to 1$ the batch is fall-dominated; when $\bar{y}_b \to 0$ it is ADL-dominated. In the implementation, this quantity is computed as \texttt{avg\_label = tf.reduce\_mean(y)} inside the compiled training step, so it reflects the actual label tensor at graph-execution time for every mini-batch.

\subsubsection{Gradient Clipping}

Before noise is added, gradients are clipped per variable tensor to a fixed $\ell_2$-norm bound $C$ to limit the sensitivity of each update. For each trainable variable $k$, the aggregated gradient $g_k = \nabla_{\theta_k}\mathcal{L}(B)$ — computed as the mean gradient over the mini-batch via \texttt{tf.GradientTape} — is clipped as:

\begin{equation}
    \tilde{g}_k \;=\; g_k \Big/\max\!\left(1,\;\frac{\|g_k\|_2}{C}\right),
    \label{eq:clipping}
\end{equation}

\noindent ensuring $\|\tilde{g}_k\|_2 \leq C$ for all variables $k$. The clipping bound is fixed at $C = 2.0$ throughout all experiments, applied via \texttt{tf.clip\_by\_norm} independently to each gradient tensor in \texttt{model.trainable\_variables}.

\subsubsection{Class-Aware Noise Scaling and Injection}

The core contribution of CA-ADP is a per-batch noise multiplier that adapts to the class composition of each mini-batch. Instead of applying a fixed noise level, the noise standard deviation is computed at every gradient step as:

\begin{equation}
    \sigma_b \;=\; \sigma_{\text{base}}\,\bigl(1 - \alpha\,\bar{y}_b\bigr),
    \label{eq:sigma_b}
\end{equation}

\noindent where $\sigma_{\text{base}}$ is the base noise multiplier, $\alpha \in [0,1)$ is the fall-adaptive factor controlling the degree of noise reduction, and $\bar{y}_b$ is the mean batch label from Eq.~\eqref{eq:batch_mean}. When the batch is fall-dominated ($\bar{y}_b \to 1$ or $\bar{y}_b=1$ ), $\sigma_b$ decreases to $\sigma_{\text{base}}(1 - \alpha)$, preserving the minority-class gradient signal. When the batch is ADL-dominated ($\bar{y}_b \to 0$ Or $\bar{y}_b=0$), $\sigma_b$ stays at $\sigma_{\text{base}}$, applying full noise to the majority class. The resulting noise range across all batches is $[\sigma_{\text{fall}},\,\sigma_{\text{ADL}}]$ = $[\sigma_{\text{base}}(1 - \alpha) ,\,\sigma_{\text{base}}]$. This adaptive noise is then injected into each clipped gradient $\tilde{g}_t$:

\begin{equation}
    \hat{g}_t \;=\; \tilde{g}_t \;+\;
    \mathcal{N}\!\left(\mathbf{0},\;\sigma_b^2\,\mathbf{I}\right),
    \label{eq:noise_injection}
\end{equation}

\noindent where $\hat{g}_t$ is the noisy gradient at step $t$ passed to the optimiser, and $\mathbf{I}$ is the identity matrix. Since the dataset
is reshuffled every epoch, $\bar{y}_b$ and therefore $\sigma_b$ vary naturally across batches without any additional scheduling logic. The final noisy gradient $\hat{g}_t$ is then used to update the model parameters:

\begin{equation}
    \theta_{t+1} \;\leftarrow\; \theta_t \;-\; \eta\cdot\hat{g}_t,
    \label{eq:update_step}
\end{equation}

\noindent where $\eta$ is the learning rate. This ensures that every parameter update incorporates both sensitivity control via gradient
clipping and privacy-preserving noise via adaptive Gaussian perturbation.

\subsection{Privacy Guarantee}

The privacy expenditure of CA-ADP is quantified using the $(\varepsilon, \delta)$-DP framework. Since the noise multiplier $\sigma_b$ varies per batch according to Eq.~\eqref{eq:sigma_b}, formal accounting is based on $\sigma_{\text{base}}$, the maximum noise value applied across all batches, which yields a conservative upper bound on the true $\varepsilon$. For a uniform sampling rate $q = |B|/N$, total gradient steps $T = E_{\text{actual}}\lfloor N/|B| \rfloor$, and failure probability $\delta$, the privacy budget can be expressed as:

\begin{equation}
    \varepsilon \;\leq\;
    \frac{q\,\sqrt{2\,T\,\ln(1/\delta)}}{\sigma_{\text{base}}}
    \label{eq:epsilon_bound}
\end{equation}

\noindent Where $N$ is the training set size, $|B|$ is the mini-batch size, and $E_{\text{actual}}$ is the number of epochs actually completed. The privacy accounting is performed using the Rényi Differential Privacy (RDP) framework~\cite{mironov2017renyi}. We used  \texttt{opacus} RDP accountant~\cite{opacus2020} to calculate the RDP. The privacy budget $\varepsilon$ is computed twice.   First, it is estimated before training as an upper bound assuming the full epoch budget. Then, after training, it is recalculated using the actual number of epochs $E_{\text{actual}}$ (the best checkpoint epoch), which gives the final reported $(\varepsilon, \delta)$ differential privacy guarantee.

\section{Proposed Hybrid 3D CNN–BiLSTM with Class-Aware Adaptive Differential Privacy}

\begin{figure}[!htb]
\begin{center}
\includegraphics[height=300px,width=450px]{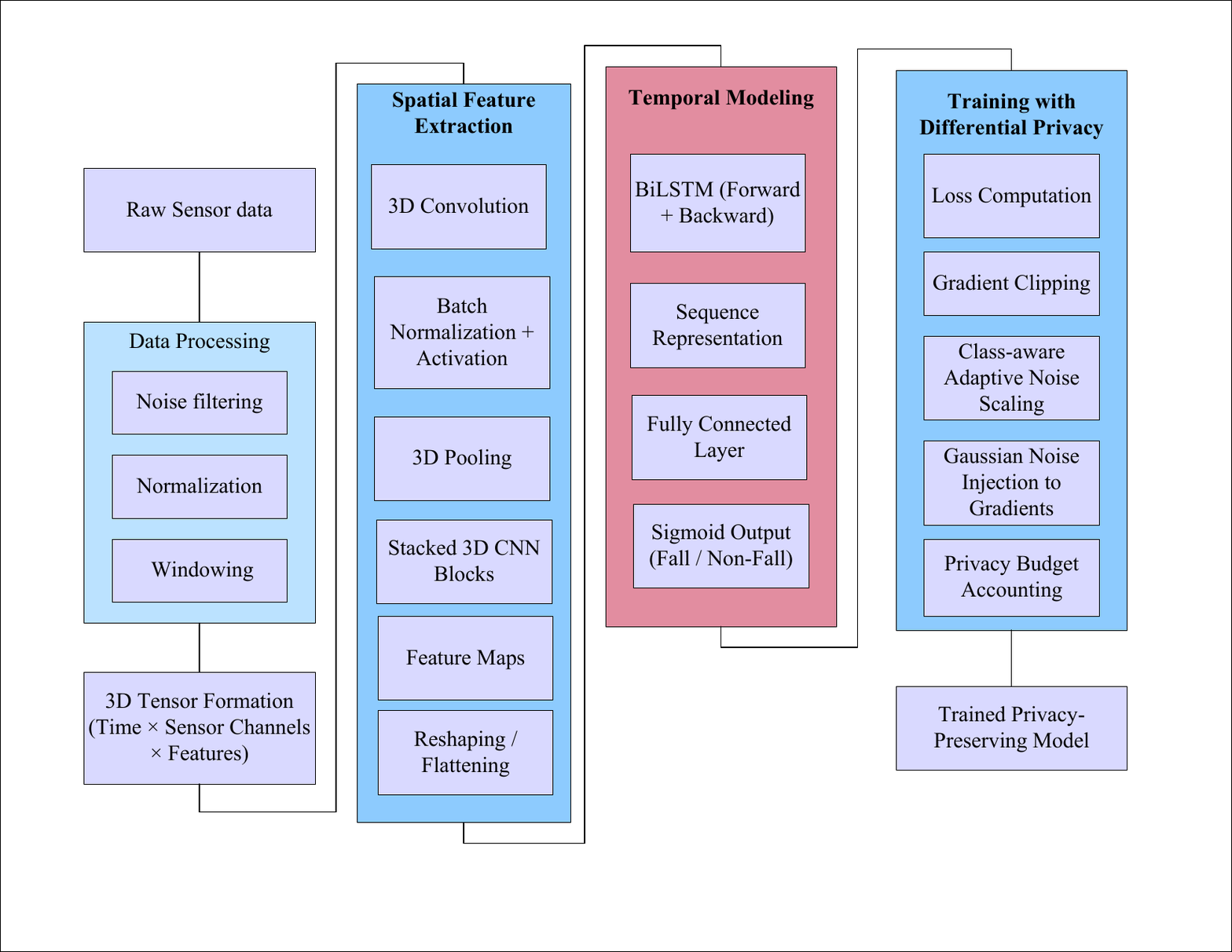}
\caption{Architecture of the proposed hybrid 3D CNN–BiLSTM model with class-aware adaptive differential privacy. The 3D CNN extracts spatiotemporal features, BiLSTM models temporal dependencies, and privacy-preserving training is achieved via class-aware gradient clipping and adaptive Gaussian noise injection. }
\label{fig:HSI_RGB_framework}
\end{center}
\end{figure}

The proposed model integrates 3D CNN, BiLSTM, and a class-aware differential privacy mechanism into a fall detection framework. The goal is to simultaneously capture both spatial patterns and temporal dynamics from sensor data, while protecting data privacy.

Initially, raw sensor signals from datasets such as SisFall, UP-Fall, and MobiAct are passed through a data processing pipeline. Each signal is segmented into fixed-length overlapping temporal windows. Each segment is converted into a structured tensor that preserves both the temporal sequence and the multi-channel nature of the data. These tensors are fed into a 3D CNN, which performs convolution across both temporal and spatial dimensions. Several stacked blocks of the framework progressively abstract higher-level feature representations from the raw sensor patterns. The extracted features are then passed to a BiLSTM network. The BiLSTM processes the sequence in both forward and backward directions. It allows the model to understand long-term temporal dependencies. This bidirectional learning helps capture contextual information before and after a potential fall event. It improves classification robustness. The BiLSTM outputs a compact representation that summarizes the full temporal activity, which is then forwarded to fully connected layers for final classification. 

To preserve privacy, a class-aware differential privacy mechanism is applied during training. Instead of using a fixed noise level the proposed approach dynamically adjusts the Gaussian noise based on the class composition of each mini-batch. Batches with a higher proportion of fall samples receive relatively lower noise and batches dominated by non-fall (ADL) samples receive higher noise. Before noise injection, all gradients are clipped using a global L2 norm, which is a fixed operation applied uniformly across all batches. This Gaussian noise is then added to each gradient before the parameter update is performed. This adaptive strategy balances privacy protection, minority class fairness, and model utility across training. This batch-wise adaptive strategy addresses the class imbalance inherent in fall detection datasets by preventing excessive noise from degrading the learning of minority class patterns. At the same time, it maintains differential privacy through controlled gradient perturbation.

 \subsection{Baseline Models}

 To evaluate the effectiveness of the proposed CA-DP based 3D CNN–BiLSTM model, we developed several baseline models and performed comparison. These baselines were designed to to examine the contribution of spatial modeling, temporal modeling, and conventional differential privacy mechanisms (DP-SGD) based hybrid modeling. First, a standalone 3D CNN model was developed to evaluate the impact of spatiotemporal feature extraction without sequential modeling. We used the same 3D CNN architecture that we used in our proposed hybrid model. Second, a pure BiLSTM model was implemented to assess the importance of temporal modeling alone. In this configuration, the preprocessed sequential sensor data were directly provided to a bidirectional LSTM network without convolutional feature extraction. We followed the same configuration of the BiLSTM model that we used in the hybrid model. Third, a hybrid 3D CNN–BiLSTM model with conventional DP-SGD was constructed to examine the effectiveness of standard differential privacy training. In this model, we used the similar architecture remains which is identical to the proposed hybrid structure. However, to enforce privacy a fixed Gaussian noise multiplier has been used across all classes. To ensure a fair comparison, all the baseline models were trained under the same experimental settings, including data preprocessing, optimizer configuration and batch size. Then, we performed a performance comparison of all the baseline models against the proposed CA-ADP–based 3D CNN–BiLSTM model.

\subsection{Evaluation Measures}
\label{subsec:evaluation_measure}

To assess the performance of the models in this study, we utilized seven standard and widely used evaluation metrics. The metics were derived from the confusion matrix entries: True Positives (TP), True Negatives (TN), False Positives (FP), and False Negatives (FN). In this fall detection study, a positive label (1) denotes a fall event, and a negative label (0) denotes an Activity of Daily Living (ADL). The description of the metrics employed in this research is given in Table \ref{table:evaluation_measures}.

\begin{table}[h!]
\caption{List of evaluation measures}
\label{table:evaluation_measures}
\begin{tabular}{  p{3cm}  p{5cm}  p{7cm}  }
 \hline
 \vspace{.05mm}\\
 Metric & Description & Equation\\
 \vspace{.05mm}\\
 \hline
 \vspace{.01mm}\\
 Accuracy & It measures the proportion of correctly prediction of the model & 
 Mathematically, Accuracy can be expressed as:
  \begin{equation}
\text{Accuracy} = \frac{TP + TN}{TP + TN + FP + FN}
\end{equation}
  \\

Specificity & It is the accurate prediction rate of non-fall activities & 
 The mathematical equation of specificity is:
 \begin{equation}
\text{Specificity} = \frac{TN}{TN + FP}
\end{equation}
  \\
  
  Precision & It is the fraction of predicted falls and actual falls of the model  & 
 Precision can be defined as:
  \begin{equation}
\text{Precision} = \frac{TP}{TP + FP}
\end{equation}
  \\

   Recall (Sensitivity) & It is the measure of actual fall events correctly detected by the model  & 
 The equation of the recall is:
 \begin{equation}
\text{Recall} = \frac{TP}{TP + FN}
\end{equation}
  \\

   F1-Score & It measures the harmonic mean of Precision and Recall  & 
The Mathematical formula of F-measure is defined below. 

 \begin{equation}
\text{F1-Score} = \frac{2 \times \text{Precision} \times \text{Recall}}
{\text{Precision} + \text{Recall}}
\end{equation}
  \\

   ROC-AUC (Area Under the Receiver Operating Characteristic Curve) &  It evaluates the model's ability to discriminate between fall and non-fall classes   &
 Mathematically, ROC-AUC can be expressed as:
  \begin{equation}
\text{ROC-AUC} = \int_{0}^{1} \text{TPR}(t)\, d\,\text{FPR}(t)
\end{equation}
where TPR and FPR denote the True Positive Rate and False Positive Rate at threshold $t$, respectively.
  \\
  
PR-AUC (Area Under the Precision-Recall Curve) & It summarizes model performance specifically with respect to the minority (fall) class across varying thresholds. Unlike ROC-AUC, PR-AUC is more 
sensitive to class imbalance and therefore serves as a stricter indicator of fall detection capability. &

The equation of the PR-AUC is:
\begin{equation}
\text{PR-AUC} = \int_{0}^{1} \text{Precision}(r)\, dr
\end{equation}
where $r$ denotes Recall.

  \\
 \hline
\end{tabular}
\end{table}

\subsection{Coding and Experimental Environment}

All experiments were carried out a workstation running Windows Server 2022 (64-bit) equipped with an Intel Xeon Silver 4214R processor clocked at 2.4 GHz, 64 GB of RAM, and 1 TB of hard disk storage. The implementation was done in Python 3.11, with Jupyter Notebook. The deep learning components were implemented using PyTorch, leveraging its \texttt{nn.Module} interface for model definition, \texttt{DataLoader} for batched data handling, and \texttt{torch.optim} for optimisation. Data preprocessing and evaluation were carried out using NumPy, SciPy, and Scikit-learn, while Pandas was used for result logging and CSV-based record keeping. Differential privacy budget estimation was performed using the Opacus \texttt{RDPAccountant} where available, with an analytical Gaussian mechanism approximation employed as a fallback. Reproducibility was ensured by fixing all random seeds across Python, NumPy, and PyTorch, and by disabling non-deterministic cuDNN operations. The complete set of code is publicly available at: \url{https://github.com/joysana1/CA-ADP}.

\section{EXPERIMENTAL RESULTS} \label{sec:experimental_result}
In order to evaluate the performance of the proposed privacy-preserving method, we conducted extensive experiments on three different datasets. The aim of these experiments is to assess fall detection performance and evaluate the privacy-preserving capability of the models. For performance evaluation, we used seven evaluation metrics and data privacy was assessed using the differential privacy parameter $\epsilon$.

\subsection{Results on Dataset-1}

\begin{figure}[!htb]
\begin{center}
\includegraphics[height=225px,width=450px]{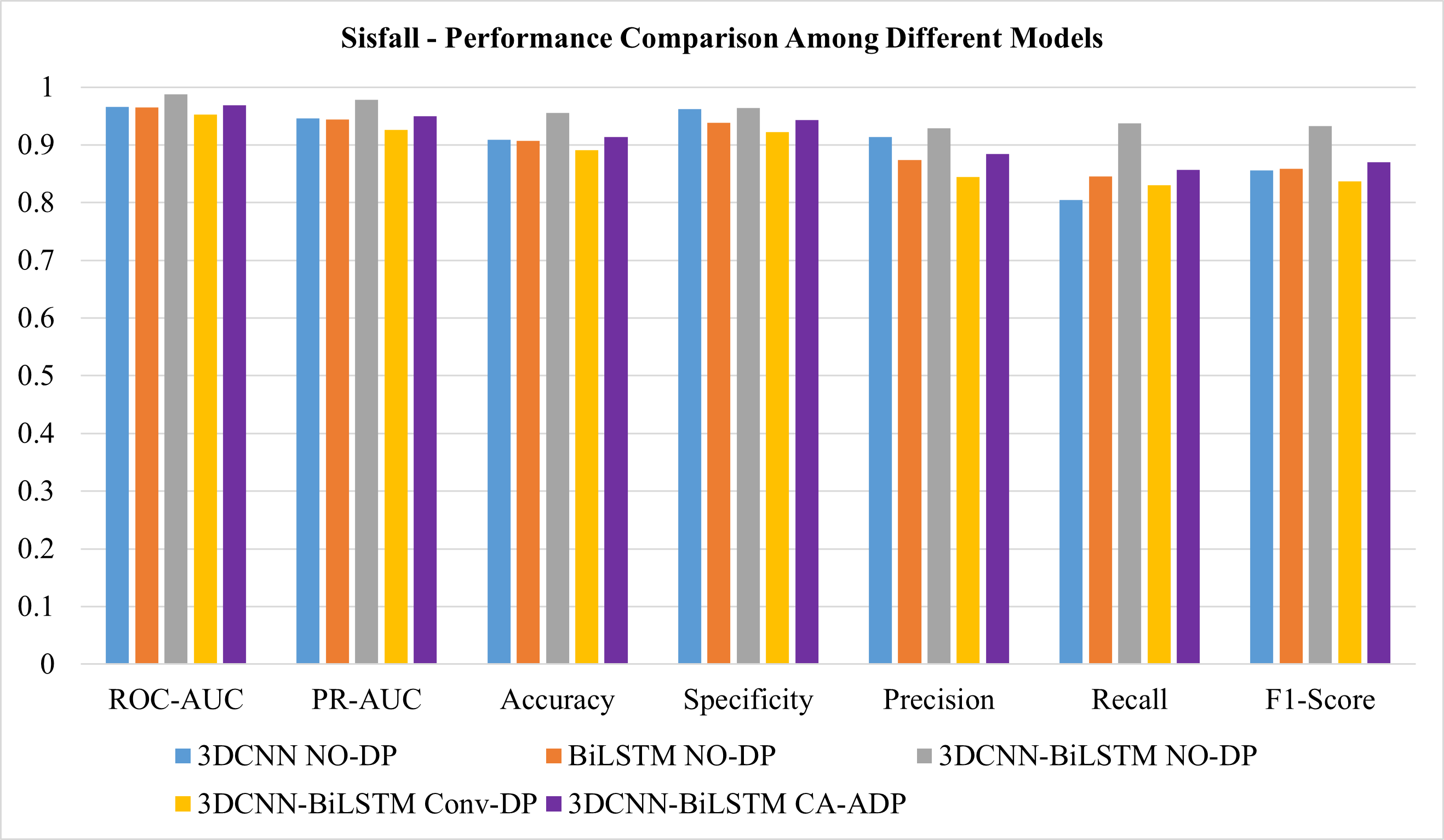}
\caption{Performance comparison among five models on the  dataset-1 across seven evaluation metrics. The 3DCNN-BiLSTM CA-ADP model achieves competitive results relative to the non-private baselines.}
\label{fig:sisfall_performance_comparison}
\end{center}
\end{figure}

\begin{figure}[!htb]
\begin{center}
\includegraphics[height=200px,width=450px]{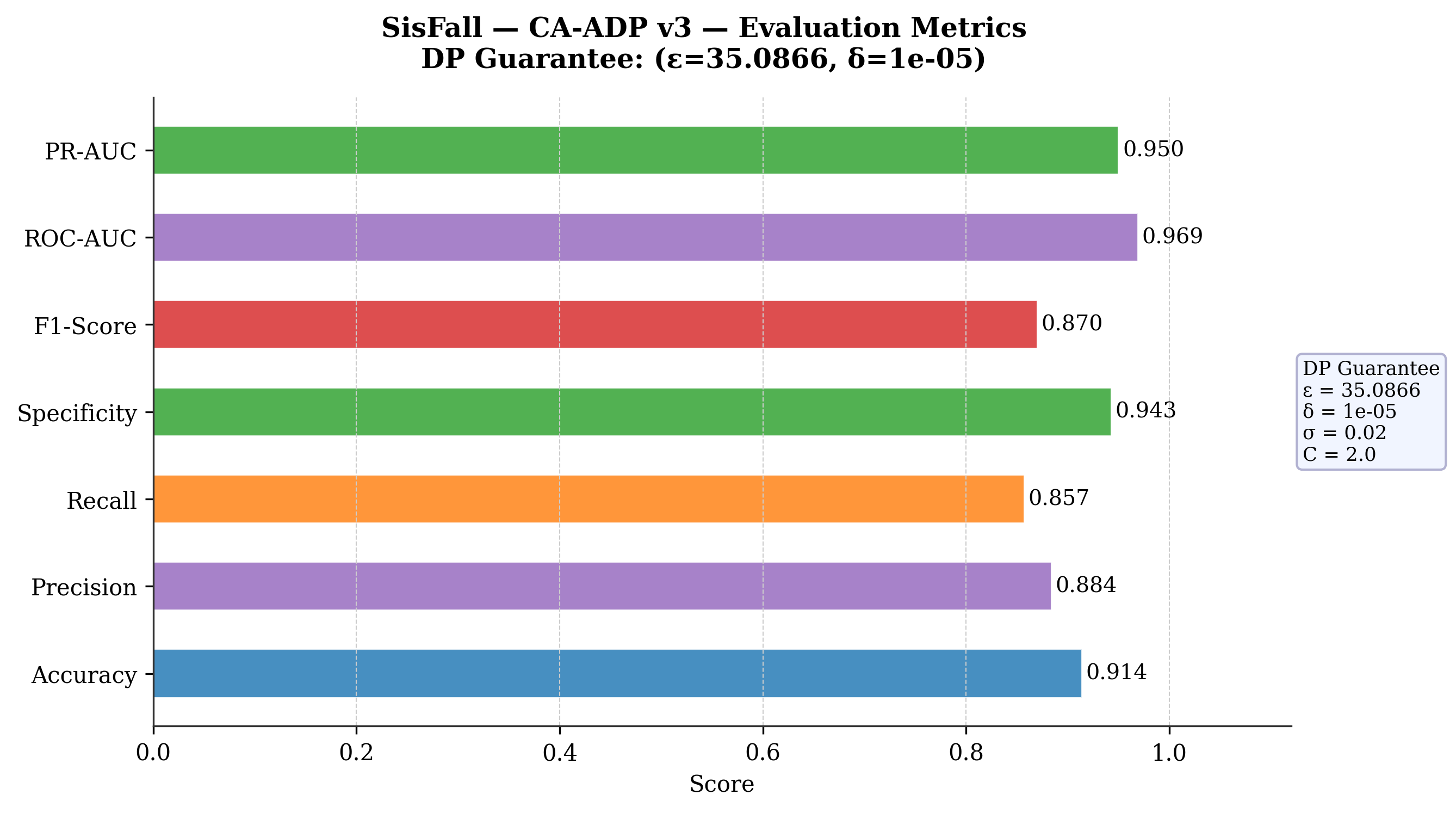}
\caption{Evaluation metrics of the 3DCNN-BiLSTM CA-ADP model on the  dataset-1 under the DP guarantee $(\varepsilon = 35.0866,\ \delta = 10^{-5},\ \sigma = 0.02,\ C = 2.0)$.}
\label{fig:sisfall_final_metrics}
\end{center}
\end{figure}

Figure ~\ref{fig:sisfall_performance_comparison} presents the experimental results of the models on the SisFall dataset, and Figures~\ref{fig:sisfall_final_metrics} shows the final scores of CA-ADP mechanism based 3DCNN-BiLSTM model in terms of all seven evaluation metrices with differential private parameters values. Among the individual models, the 3DCNN and BiLSTM achieved comparable baseline performance, with ROC-AUC scores of 0.966 and 0.965 and F1-scores of 0.856 and 0.859, respectively. The hybrid 3DCNN-BiLSTM model without any differential privacy constraint (NO-DP) achieved the strongest overall performance, with a ROC-AUC of 0.988, PR-AUC of 0.978, accuracy of 0.955, and an F1-score of 0.933 (macro-average F1-score of 0.95 and weighted-average F1-score of 0.96), confirming the benefit of jointly modeling spatial and temporal features. When differential privacy was applied, the Conv-DP (DP-SGD) configuration exhibited a notable utility drop, yielding a ROC-AUC of 0.953 and F1-score of 0.837 (macro avg F1 is 0.87 and weighted avg F1 is 0.88), reflecting the well-known trade-off between strict noise injection and model accuracy. In contrast, the proposed CA-ADP mechanism achieved a ROC-AUC of 0.969, PR-AUC of 0.950, accuracy of 0.914, specificity of 0.943, precision of 0.884, recall of 0.857, and F1-score of 0.870 under a formal DP guarantee of $(\varepsilon = 35.0866,\ \delta = 1 \times 10^{-5})$, as shown in Figure~\ref{fig:sisfall_final_metrics}.

\subsection{Results on Dataset-2} \label{sec:exp_dataset2}
\begin{figure}[!htb]
\begin{center}
\includegraphics[height=225px,width=450px]{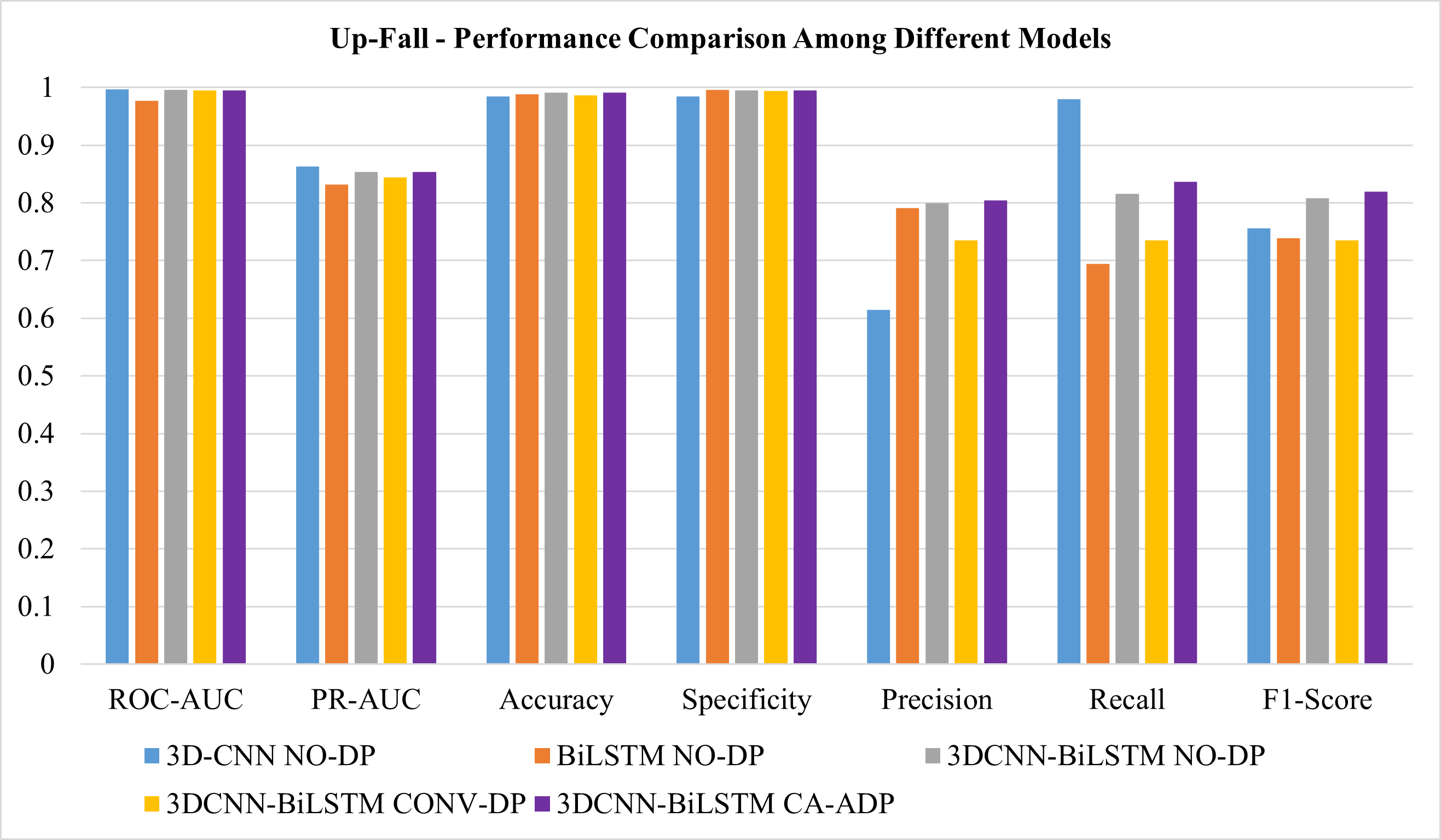}
\caption{Performance comparison among five models on the dataset-3 across seven evaluation metrics. The 3DCNN-BiLSTM CA-ADP model achieves competitive results relative to the non-private baselines.}
\label{fig:upfall_performance_comparison}
\end{center}
\end{figure}

\begin{figure}[!htb]
\begin{center}
\includegraphics[height=200px,width=450px]{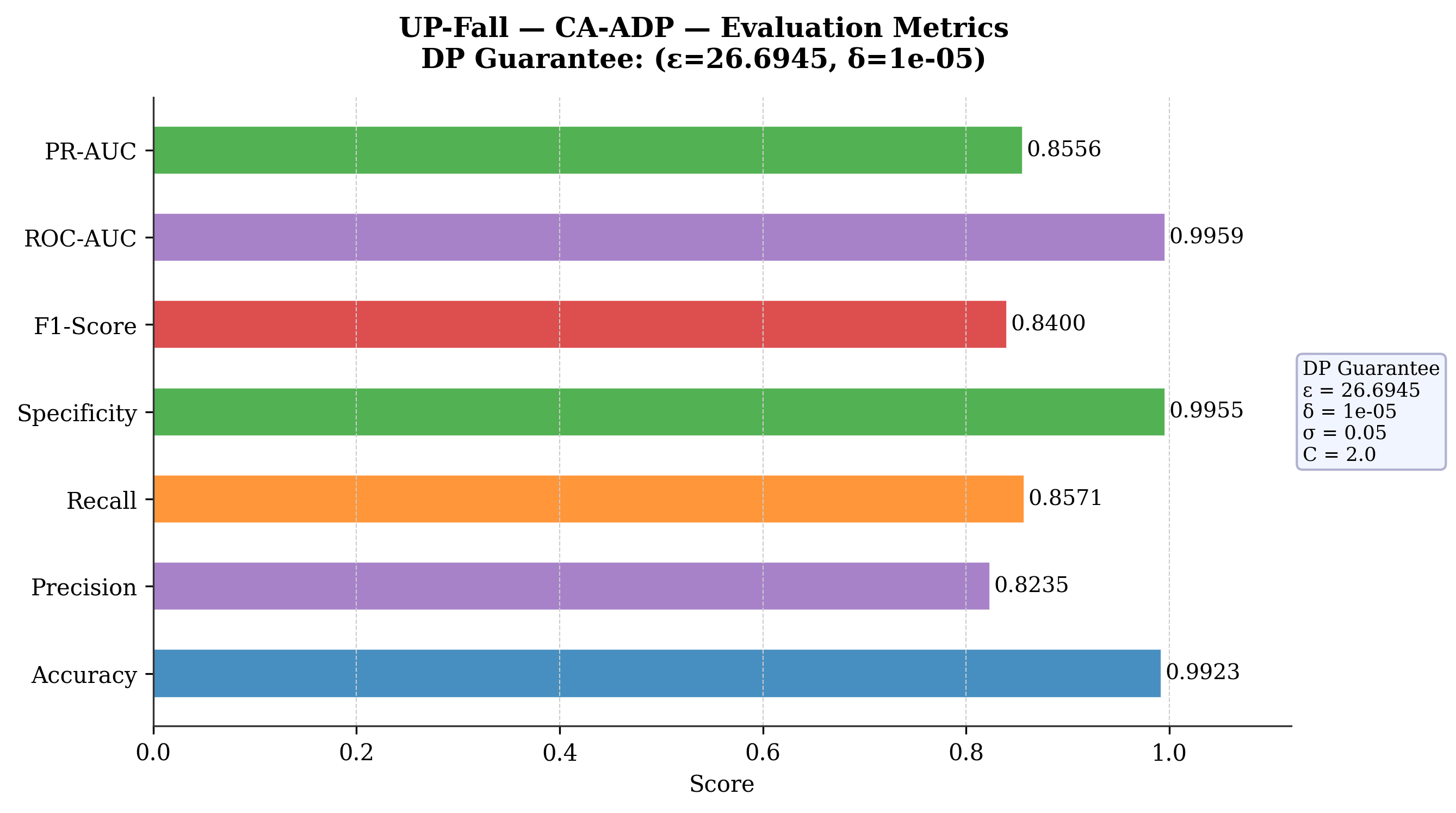}
\caption{Evaluation metrics of the 3DCNN-BiLSTM CA-ADP model on the  dataset-2 under the DP guarantee $(\varepsilon = 26.6945,\ \delta = 10^{-5},\ \sigma = 0.05,\ C = 2.0)$.}
\label{fig:upfall_final_metrics}
\end{center}
\end{figure}

For the Up-Fall dataset, the performance comparison among the models are presented in the Figure~\ref{fig:upfall_performance_comparison} across the all evaluation metrics. The evaluation scores of the propsoed CA-ADP based hybrid 3DCNN-BiLSTM model with differentail privacy parameters values are illustrated in the Figure~\ref{fig:upfall_final_metrics}. Among the baseline models, the standalone 3D-CNN achieved the highest ROC-AUC (0.99) and recall (0.980), but its low precision  (0.615) yielded a modest F1-Score of 0.756, indicating a tendency toward false positives, while the BiLSTM exhibited more conservative behavior with higher precision (0.791) but lower recall (0.694) and an F1-Score of 0.739. The hybrid 3DCNN-BiLSTM without DP outperformed both baselines with an accuracy of 0.99 and an F1-Score of 0.81, confirming the complementary benefit of joint spatiotemporal and sequential modeling, whereas applying CONV-DP (DP-SGD) degraded performance to an F1-Score of 0.735. The proposed CA-ADP achieved the best overall balance with an F1-Score of 0.820 and better recall of 0.837 than the CONV-DP based hybrid model and pure 3DCNN-BiLSTM model. 

\subsection{Results on dataset-3} \label{sec:exp_dataset3}

\begin{figure}[!htb]
\begin{center}
\includegraphics[height=225px,width=450px]{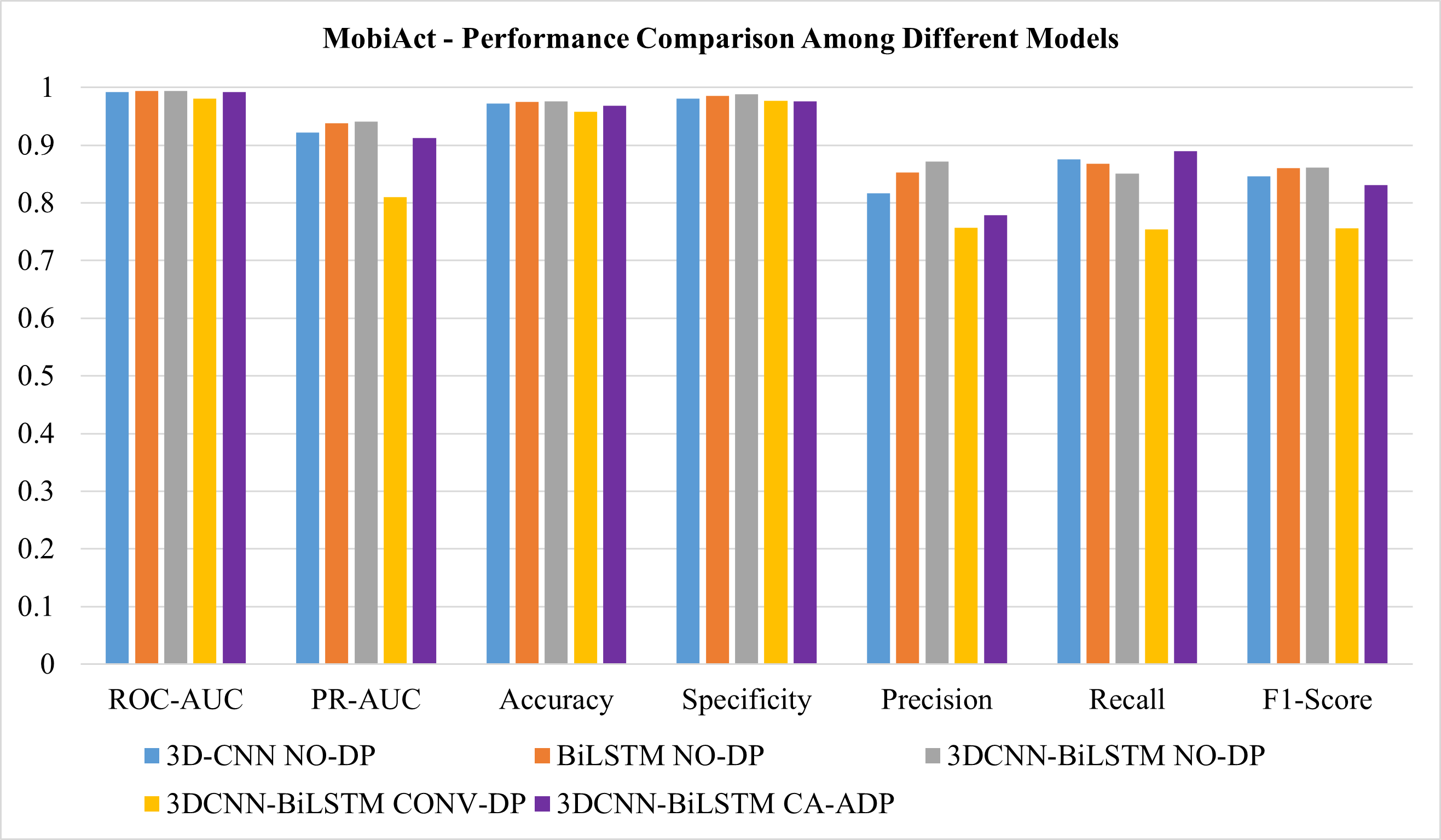}
\caption{Performance comparison among five models on the dataset-3 across seven evaluation metrics.}
\label{fig:mobiAct_fall_performance_comparison}
\end{center}
\end{figure}

\begin{figure}[!htb]
\begin{center}
\includegraphics[height=200px,width=450px]{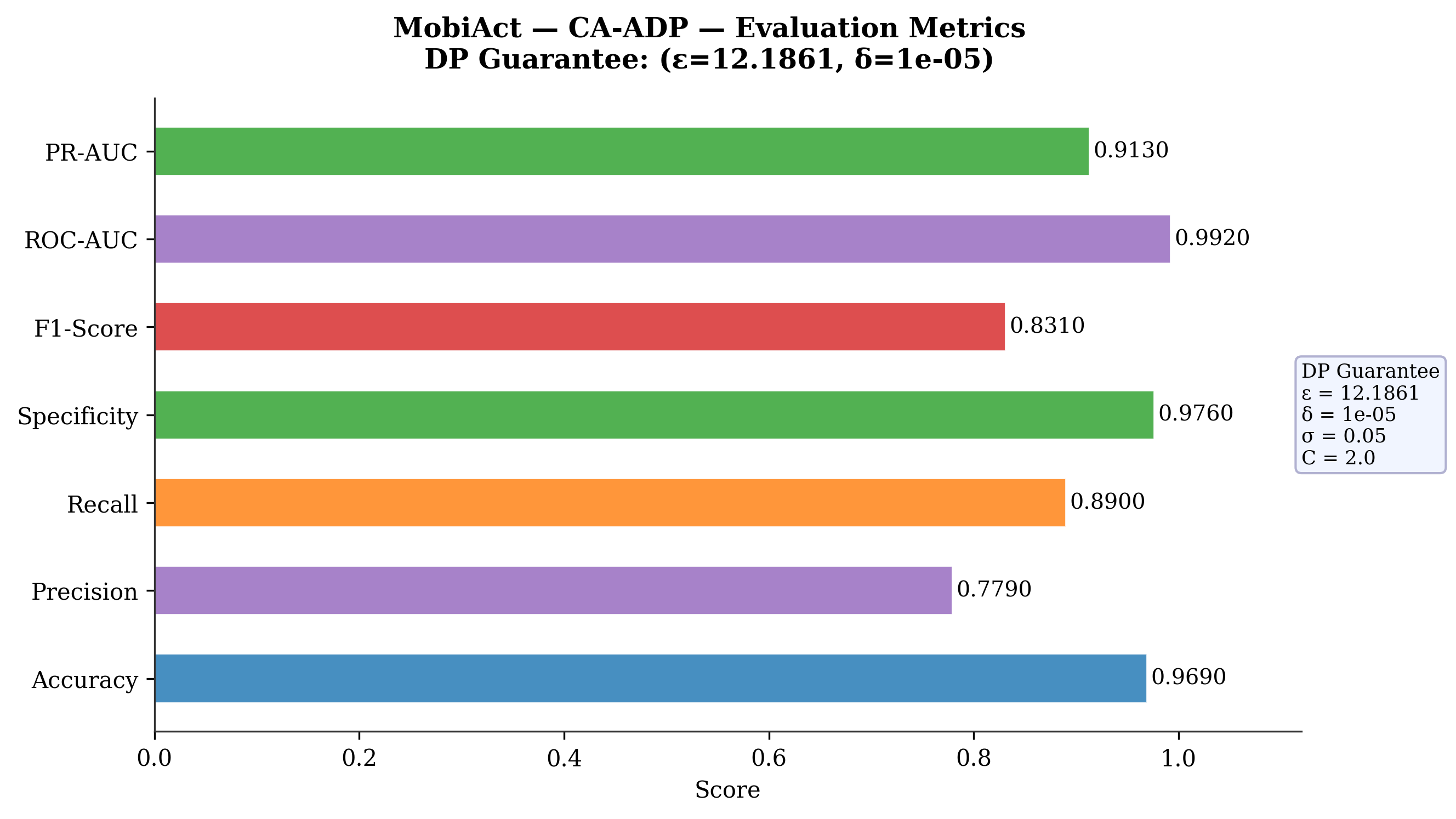}
\caption{Evaluation metrics of the 3DCNN-BiLSTM CA-ADP model on the  dataset-3 under the DP guarantee $(\varepsilon = 12.1861,\ \delta = 10^{-5},\ \sigma = 0.05,\ C = 2.0)$.}
\label{fig:mobiact_final_metrics}
\end{center}
\end{figure}

Figure~\ref{fig:mobiAct_fall_performance_comparison}
presents the performance of all evaluated models on the MobiAct dataset. Figure ~\ref{fig:mobiact_final_metrics} shows the values of differential privacy parameters with evaluation metrics. Among the models the hybrid 3DCNN-BiLSTM (No DP) model achieved the best overall performance. It obtained the highest Accuracy (0.976) and F1-Score (0.861), along with strong ROC-AUC (0.994) and PR-AUC (0.941). This indicates that combining spatial feature extraction (3D-CNN) with temporal modeling (BiLSTM) leads to superior classification capability. On the other hand, the CONV-DP–based 3D CNN–BiLSTM model shows a noticeable performance drop across all metrics. The PR-AUC decreased to 0.81, and the F1-score dropped to 0.756. This indicates that the CONV-DP–based model struggles to handle the class imbalance issue. In contrast, the proposed CA-ADP method demonstrates a better privacy-utility trade-off. Although there is a slight decline compared to the NO-DP setting, it staill maintains strong performance. It shows ROC-AUC of 0.992, Accuracy of 0.969, and F1-Score of 0.831. Notably, the CA-ADP–based model achieves the highest recall (0.89) among all models, indicating improved sensitivity in detecting positive cases. This improvement is critical for many real-world applications.

\section{Statistical Performance Test} \label{sec:statistical_test}

To see whether the performance gains of CA-ADP over CONV-DP were statistically significant or not, we performed Wilcoxon signed-rank test (WSRT)~\cite{JDK_2025_PPCCP}\cite{Janez2006}. As a non-parametric paired test, WSRT does not assume normality. All the seven evaluation metrics have been used as paired observations for each dataset. The significance level was set to $\alpha = 0.05$, and a Bonferroni correction ($\alpha_{\text{corrected}} = 0.05/3 \approx 0.0167$) was applied to account for multiple comparisons.

As shown in Table~\ref{tab:wilcoxon}, CA-ADP achieved significant improvements over the CONV-DP. The test produces $W = 0$ ($p = 0.0078$) on the SisFall dataset, $W = 0$ ($p = 0.0156$) on the Up-Fall dataset, and $W = 1$ ($p = 0.0156$) on the MobiAct dataset. The zero W-statistic of SisFall and Up-Fall suggests that CA-ADP outperformed CONV-DP on every single metric. This finding was further supported by the pooled test over all 21 metric pairs (a combination of all three datasets), which yielded $W = 1.5$ and $p = 0.0001$. All these outputs confirm that the superiority of CA-ADP is not due to chance.

The per-metric analysis, summarized in Table~\ref{tab:metric_diff}, provides a clearer picture of where these improvements come from. CA-ADP consistently improved across all metrics on all datasets, with an exception in terms of Specificity on MobiAct ($-0.0010$), which is negligible. The most noticeable improvements are seen in Recall ($+0.0883$) and F1-score ($+0.0643$). This suggests that CA-ADP significantly enhances the model’s ability to detect fall events. It is a very important factor in real-world settings because missing a fall can be far more serious than raising a false alarm. The gains in PR-AUC ($+0.0457$) and Precision ($+0.0437$) further indicate that the model produces more reliable predictions. It reduces false positives compared to CONV-DP. These findings show that the improvements offered by CA-ADP are not only statistically significant but also practically meaningful.

\begin{table}[ht]
\centering
\caption{Wilcoxon Signed-Rank Test Results Comparing 3DCNN-BiLSTM-CONV-DP and 3DCNN-BiLSTM-CA-ADP Across Three Datasets. Seven evaluation metrics per dataset serve as paired observations. Bonferroni-corrected significance threshold: $\alpha_{\text{corrected}} = 0.05/3 \approx 0.0167$.}
\label{tab:wilcoxon}
\resizebox{\columnwidth}{!}{%
\begin{tabular}{lccccc}
\toprule
\textbf{Dataset} &
\textbf{W-statistic} &
\textbf{p-value} &
\textbf{Sig. ($\alpha$=0.05)} &
\textbf{Sig. (Bonferroni)} &
\textbf{Mean Diff.} \\
\midrule
SisFall         & 0.0000 & 0.0078 & \checkmark & \checkmark & $+0.0263$ \\
Up-Fall         & 0.0000 & 0.0156 & \checkmark & \checkmark & $+0.0387$ \\
MobiAct         & 1.0000 & 0.0156 & \checkmark & \checkmark & $+0.0510$ \\
\midrule
\textbf{Pooled} & \textbf{1.5000} & \textbf{0.0001} 
                & \checkmark & \checkmark & $+0.0387$ \\
\bottomrule
\end{tabular}%
}
\smallskip
\begin{minipage}{\columnwidth}
\footnotesize
\textit{Note:} One-sided Wilcoxon signed-rank test with $H_1$: CONV-DP $<$ CA-ADP. Seven metrics (ROC-AUC, PR-AUC, Accuracy, Specificity, Precision, Recall, F1-Score) used as paired observations 
per dataset. Pooled test uses all 21 pairs across three datasets. \checkmark~denotes statistical significance.
\end{minipage}
\end{table}

\begin{table}[ht]
\centering
\caption{Per-Metric Performance Differences (CA-ADP $-$ CONV-DP) Across Three Datasets. Positive values indicate improvement by CA-ADP.}
\label{tab:metric_diff}
\begin{tabular}{lcccc}
\toprule
\textbf{Metric} &
\textbf{SisFall} &
\textbf{Up-Fall} &
\textbf{MobiAct} &
\textbf{Mean} \\
\midrule
ROC-AUC     & $+0.0160$ & $+0.0000$ & $+0.0110$ & $+0.0090$ \\
PR-AUC      & $+0.0240$ & $+0.0100$ & $+0.1030$ & $+0.0457$ \\
Accuracy    & $+0.0230$ & $+0.0040$ & $+0.0110$ & $+0.0127$ \\
Specificity & $+0.0210$ & $+0.0010$ & $-0.0010$ & $+0.0070$ \\
Precision   & $+0.0400$ & $+0.0690$ & $+0.0220$ & $+0.0437$ \\
Recall      & $+0.0270$ & $+0.1020$ & $+0.1360$ & $+0.0883$ \\
F1-Score    & $+0.0330$ & $+0.0850$ & $+0.0750$ & $+0.0643$ \\
\midrule
\textbf{Mean} & $+0.0263$ & $+0.0387$ & $+0.0510$ & $+0.0387$ \\
\bottomrule
\end{tabular}
\smallskip
\begin{minipage}{\columnwidth}
\footnotesize
\textit{Note:} The only negative value (Specificity on MobiAct, $-0.0010$) is negligibly small and does not affect the overall conclusion. The largest consistent gains are observed in Recall 
($+0.088$ mean) and F1-Score ($+0.064$ mean).
\end{minipage}
\end{table}

\section{Privacy-Utility Trade-off Analysis}
\label{sec:trade-off_analysis}

This section examines the privacy-utility trade-off inherent in our Class-Aware Adaptive Differential Privacy (CA-ADP) mechanism across the three benchmark datasets: SisFall, UP-Fall, and MobiAct. For each dataset, we analyze how F1-score varies for different privacy budget $\varepsilon$, and how the noise multiplier $\sigma$ changes the resulting $\varepsilon$ values under the Differential Privacy accountant.

\begin{figure}[!htb]
\begin{center}
\includegraphics[height=200px,width=450px]{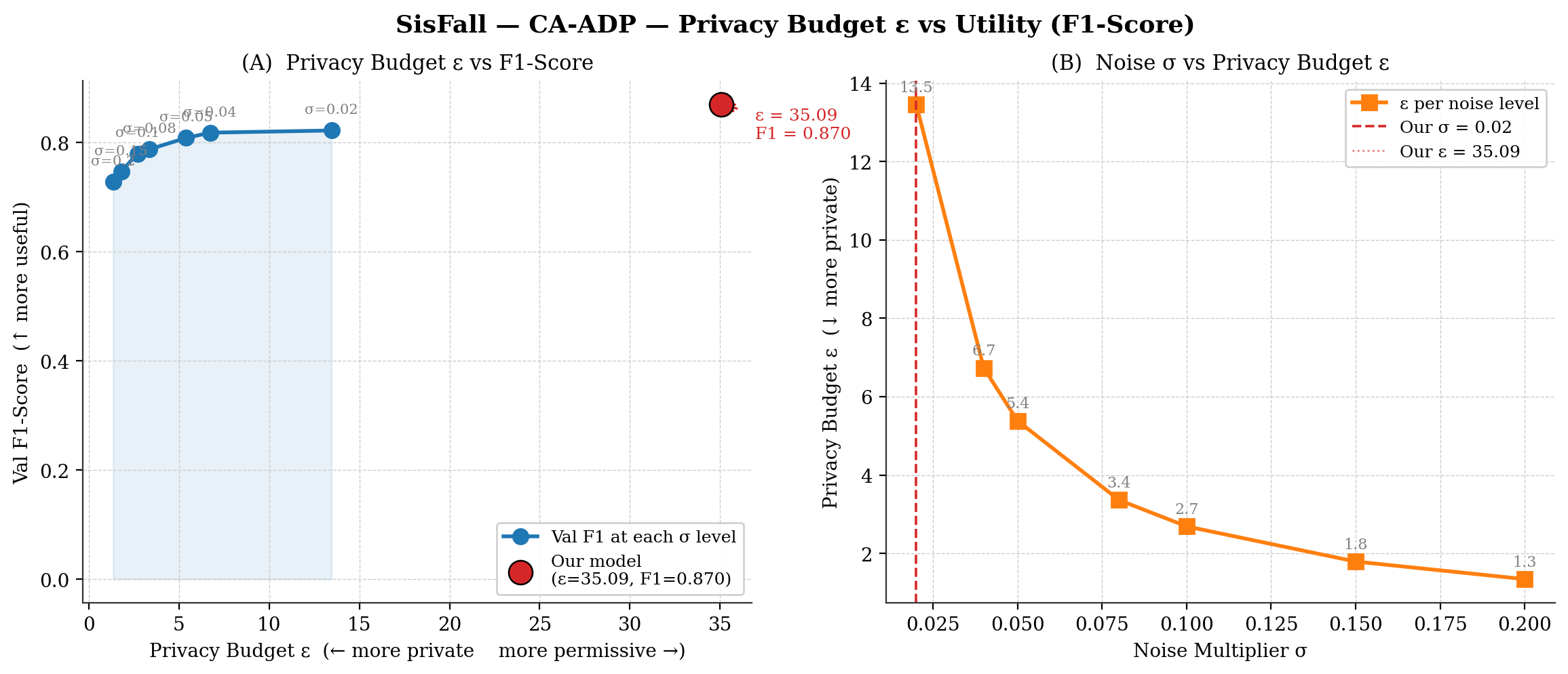}
\caption{Privacy–utility trade-off for the SisFall dataset under CA-ADP. Subplot~(A) shows the validation F1-score as a function of the privacy budget $\varepsilon$ for noise multipliers $\sigma \in [0.02, 0.20]$. The selected operating point (red marker) corresponds to epoch 47, achieving $\varepsilon = 35.09$ and an F1-score of 0.870. Subplot~(B) shows the monotonic relationship between $\sigma$ and $\varepsilon$, computed using the Rényi differential privacy accountant at $10$ training epochs, with the chosen $\sigma = 0.02$ indicated by the red dashed line.}
\label{fig:sisfall_epsilon_tradeoff}
\end{center}
\end{figure}

\begin{figure}[!htb]
\begin{center}
\includegraphics[height=200px,width=450px]{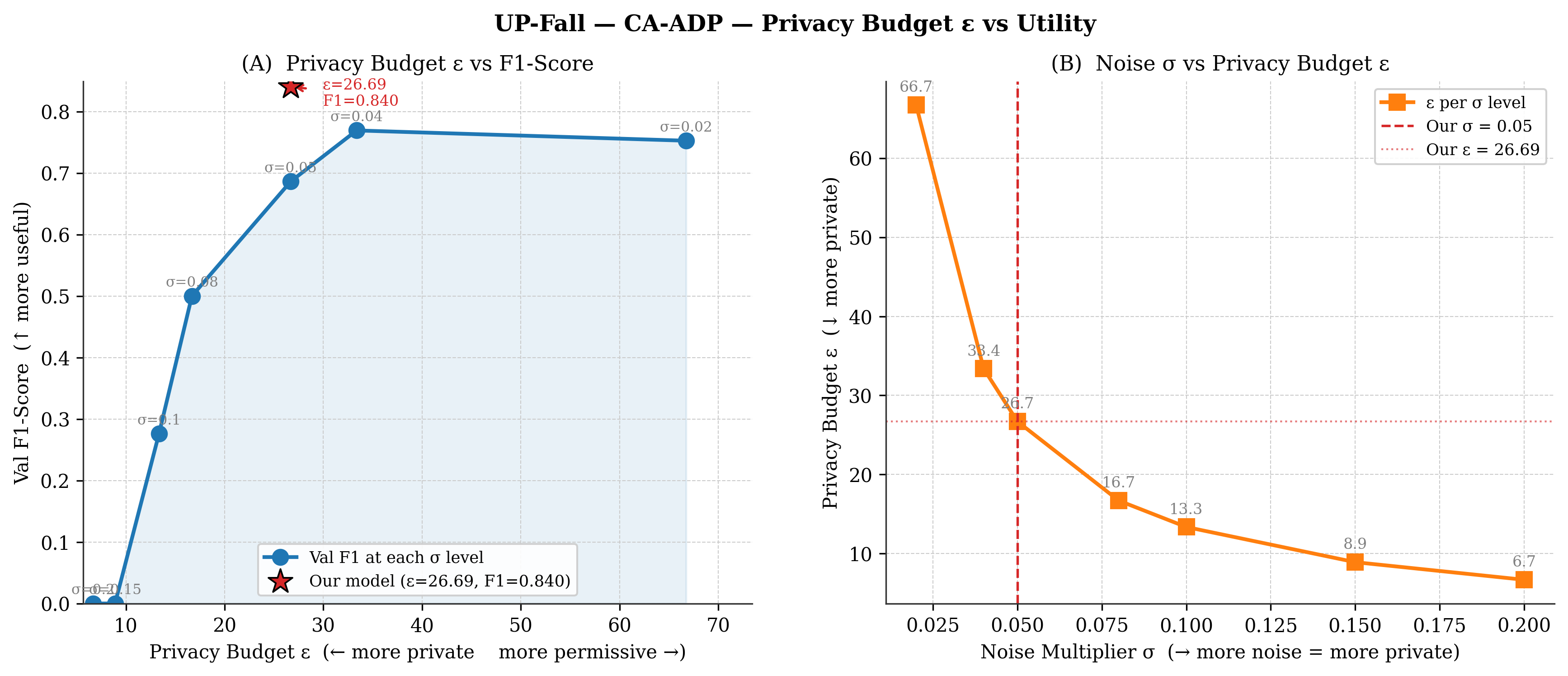}
\caption{Privacy–utility trade-off for the UP-Fall dataset under CA-ADP. Subplot~(A) presents the validation F1-score as a function of the privacy budget $\varepsilon$ for noise multipliers $\sigma \in [0.02, 0.20]$. The selected operating point (red star) corresponds to epoch 20, achieving $\varepsilon = 26.69$ and an F1-score of 0.840. Subplot~(B) illustrates the relationship between $\sigma$ and $\varepsilon$, computed using the Rényi differential privacy accountant at $10$ training epochs, highlighting the steep trade-off curve. The chosen $\sigma = 0.05$ and corresponding $\varepsilon = 26.69$ are indicated by red dashed lines.}
\label{fig:upfall_epsilon_tradeoff}
\end{center}
\end{figure}

\begin{figure}[!htb]
\begin{center}
\includegraphics[height=200px,width=450px]{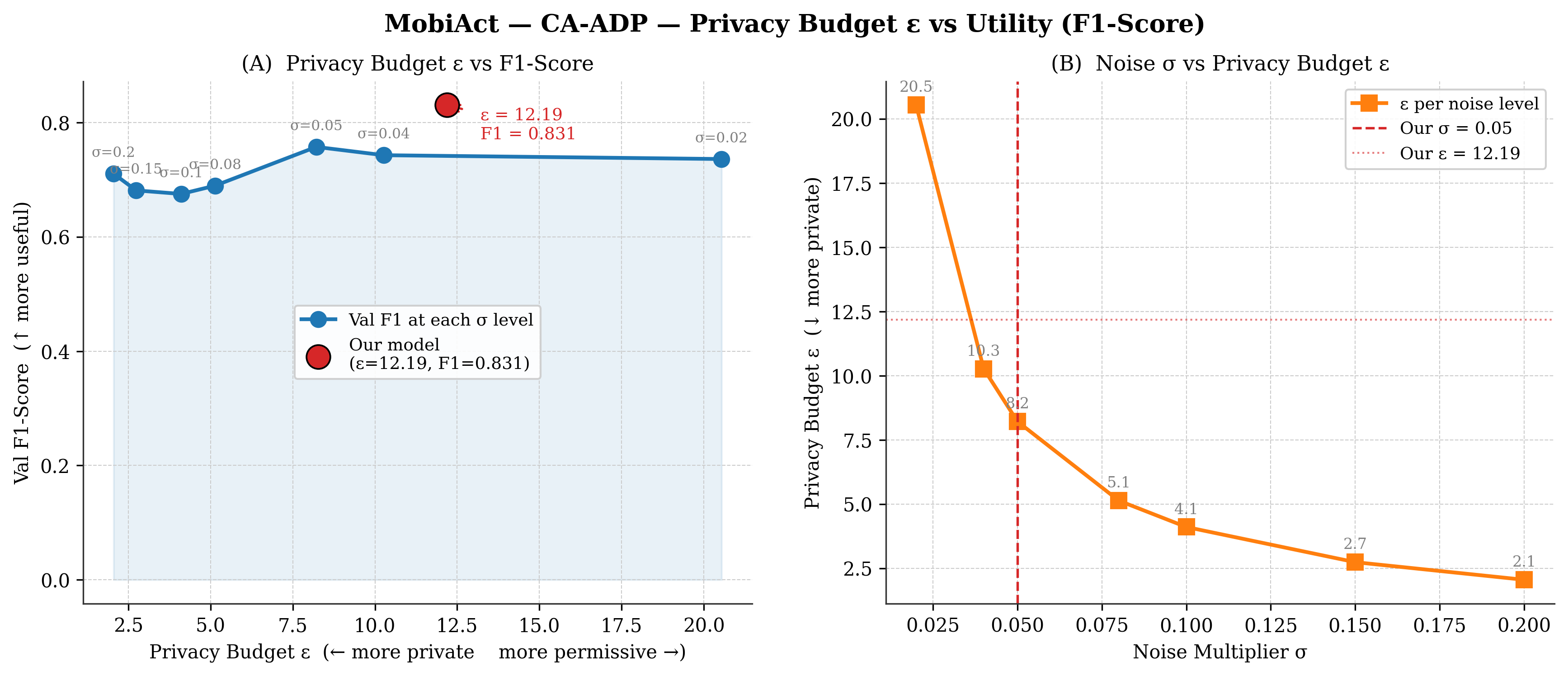}
\caption{Privacy–utility trade-off for the MobiAct dataset under CA-ADP. Subplot~(A) presents the validation F1-score as a function of the privacy budget $\varepsilon$ for noise multipliers $\sigma \in [0.02, 0.20]$. The selected operating point (red marker) corresponds to epoch 22, achieving $\varepsilon = 12.19$ and an F1-score of 0.825. Subplot~(B) illustrates the relationship between $\sigma$ and $\varepsilon$, computed using the Rényi differential privacy accountant at $10$ training epochs,, showing a monotonic decrease in $\varepsilon$ as $\sigma$ increases. The chosen $\sigma = 0.05$ is indicated by the red dashed line.}
\label{fig:mobiact_epsilon_tradeoff}
\end{center}
\end{figure}

The privacy budget vs. utility trade-off analysis for the SisFall dataset is presented in Figure~\ref{fig:sisfall_epsilon_tradeoff}. Subfigure~(B) illustrates the theoretical relationship between the noise multiplier $\sigma$. The privacy budget $\varepsilon$, computed via the Rényi Differential Privacy (RDP) accountant at a fixed reference of $10$ training epochs. As the noise multiplier $\sigma$ increases from $0.02$ to $0.20$, the privacy budget $\varepsilon$ decreases from $13.5$ to $1.3$. It confirms the inverse relationship between noise level and privacy cost. Subfigure~(A) shows how the validation F1-score changes across these $\sigma$ values. The F1-score decreases gradually from $0.82$ at $\sigma = 0.02$ to $0.72$ at $\sigma = 0.20$. Notably, the CA-ADP operating point $(\varepsilon = 35.09,\ \text{F1} = 0.870)$ lies beyond the sweep range of the conventional trade-off curve. This indicates that the adaptive class-aware noise strategy has the potential to balance between privacy and utility.

To assess the privacy-utility trade-off of CA-ADP method on the UP-Fall dataset, we tested the model across different levels of Gaussian noise ($\sigma$). The results are illustrated in Figure~\ref{fig:upfall_epsilon_tradeoff}. Subplot~(A) shows that model performance drops sharply in high-privacy settings. At $\sigma = 0.15$ and $\sigma = 0.20$, the F1-score falls to around $0.01$–$0.02$. This indicates that at higher $\sigma$ the model nearly loses its ability to make meaningful predictions. As the noise level decreases and the privacy budget $\varepsilon$ increases, the model gradually recovers. The graph shows that the F1-score improves to $0.50$ at $\sigma = 0.08$ and reaches $0.77$ at $\sigma = 0.04$. Based on this trend, the selected configuration at $\sigma = 0.05$ achieves a good balance. The values $\varepsilon = 26.69$ and an F1-score of $0.840$ represent the optimal operating point on the privacy–utility curve. Subplot~(B) illustrates the theoretical relationship between the noise level and the privacy budget $\varepsilon$. The curve is quite steep. The $\varepsilon$ starting as high as $66.7$ at $\sigma = 0.02$ and dropping quickly to $6.7$ at $\sigma = 0.20$. This indicates the proposed method effectively manage the privacy–utility trade-off.

To evaluate the privacy-utility trade-off of CA-ADP on the MobiAct dataset, we trained the model using different noise multipliers $\sigma$, as shown in Figure~\ref{fig:mobiact_epsilon_tradeoff}. The F1-score rises from around $0.72$ at $\sigma = 0.20$ ($\varepsilon \approx 2.1$) to a peak of $0.825$ at $\varepsilon = 12.19$ ($\sigma = 0.05$) and  stays close to this level even when the privacy budget is further relaxed to $\varepsilon \approx 20.5$ at $\sigma = 0.02$. The chosen operating point at $\sigma = 0.05$ gives $\varepsilon = 12.19$ with an F1-score of $0.831$, which represents a good balance between privacy and performance. Subplot~(B) further shows the expected trend: as $\sigma$ increases, $\varepsilon$ decreases sharply.  $\varepsilon$ drops from about $20.5$ at $\sigma = 0.02$ to $2.1$ at $\sigma = 0.20$, with the selected point highlighted at $\sigma = 0.05$. These findings indicate that the CA-ADP method maintains robust and stable fall detection performance under differential privacy constraints, with minimal degradation.

\section{Differential Privacy Guarantee Comparison} \label{DP-guarntee-com}
To assess the privacy cost of the proposed approach, we compare the privacy budgets achieved by the CA-ADP mechanism with those of the CONV-DP baseline across all three datasets. The privacy parameter is fixed at $\delta = 10^{-5}$ for all experiments.  The resulting privacy budgets ($\varepsilon$) are presented in Table~\ref{tab:dp_comparison}. In this context, a smaller $\varepsilon$ reflects stronger privacy protection.

\begin{table}[h]
\centering
\caption{Comparison of privacy budgets ($\varepsilon$) between CA-ADP and CONV-DP at $\delta = 10^{-5}$.}
\label{tab:dp_comparison}
\begin{tabular}{lcc}
\toprule
\textbf{Dataset} & \textbf{CA-ADP} $(\varepsilon)$ & \textbf{CONV-DP(DP-SGD)} $(\varepsilon)$ \\
\midrule
SisFall  & 35.0866  & 65.641       \\
UpFall   & 26.6945  & 53.3891      \\
MobiAct  & 12.1861  & 11.9925 \\
\bottomrule
\end{tabular}
\end{table}

CA-ADP achieves lower privacy budgets than CONV-DP (DP-SGD) on two of the three datasets, indicating stronger formal privacy protection in those cases. On the SisFall dataset, CA-ADP shows $\varepsilon = 35.0866$ compared to $\varepsilon = 65.641$ for DP-SGD, representing an improvement of approximately $46.5\%$. Similarly, on the UpFall dataset, CA-ADP reduces the privacy budget to $\varepsilon = 26.6945$ from $\varepsilon = 53.3891$, a reduction of roughly $50\%$.  In contrast, the MobiAct dataset exhibits a slightly different behavior. Here, CA-ADP results in a marginally higher privacy budget ($\varepsilon = 12.1861$) compared to CONV-DP ($\varepsilon = 11.9925$), although the difference remains below $2\%$. This minor variation may be attributed to differences in the underlying data distribution. These findings demonstrate that CA-ADP achieves competitive or superior privacy guarantees across diverse datasets.

\section{Comparison with Other studies} \label{sec:comp_other_study}
We compare the performance of the proposed models with prior studies evaluated on the same datasets. Tables~\ref{table:PER_COMP_pre_study_dataset_1}, \ref{table:PER_COMP_pre_study_dataset_2}, and \ref{table:PER_COMP_pre_study_dataset_3} present dataset specific comparisons for SisFall, UP-Fall, and MobiAct, respectively. For Dataset-1 and Dataset-2, performance comparisons are reported in terms of accuracy and F1-score. Following the evaluation criteria used in prior work~\cite{AL_QANESS_2024}, Macro-average F1 and Weighted-average F1 scores are used for Dataset-3. Some of the existing studies included in this comparison employed $K$-fold cross-validation approach. It is important to note that $K$-fold cross-validation may provide biased estimation~\cite{cawley2010overfitting}. Despite this methodological advantage of these studies, the proposed CA-ADP based 3D-CNN-BiLSTM framework achieves competitive or superior prediction performance across all three datasets.  In addition, our proposed CA-ADP-based models offer differential privacy guarantees. On the other hand, previous studies did not consider the data privacy issues.

\begin{table}[hbt!]
\begin{center}
\caption{Performance comparison between this study and previous studies in~\cite{SisFall_2017} and~\cite{Ahn_2019} using Dataset-1 (SisFall).}
\label{table:PER_COMP_pre_study_dataset_1}
\renewcommand{\arraystretch}{1.5}
\begin{tabular}{p{8cm} p{2.0cm} p{2.0cm} p{2.0cm}}
\hline
Study & Accuracy & F-Measure & Data Privacy \\
\hline
CA-ADP-3D-CNN-BiLSTM (proposed) & 0.914 & 0.870 & Yes \\
3D-CNN-BiLSTM (proposed, no DP) & 0.955 & 0.933 & No  \\
\cite{SisFall_2017}              & $<$0.90 (avg.) & ---  & No  \\
\cite{Ahn_2019}                  & 0.903 & ---  & No  \\
\hline
\end{tabular}
\end{center}
\end{table}

\begin{table}[hbt!]
\begin{center}
\caption{Performance comparison between this study and previous studies in~\cite{Mostafa_2024} using Dataset-2 (UP-Fall).}
\label{table:PER_COMP_pre_study_dataset_2}
\renewcommand{\arraystretch}{1.5}
\begin{tabular}{p{8cm} p{2.0cm} p{2.0cm} p{2.0cm}}
\hline
Study & Accuracy & F-Measure & Data Privacy \\
\hline
CA-ADP-3D-CNN-BiLSTM (proposed) & 0.995 & 0.820 & Yes \\
3D-CNN-BiLSTM (proposed, no DP) & 0.996 & 0.808 & No  \\
\cite{Mostafa_2024}              & 0.930 & ---  & No  \\
\hline
\end{tabular}
\end{center}
\end{table}

\begin{table}[hbt!]
\begin{center}
\caption{Performance comparison between this study and previous studies in~\cite{AL_QANESS_2024} using Dataset-3 (MobiAct).}
\label{table:PER_COMP_pre_study_dataset_3}
\renewcommand{\arraystretch}{1.5}
\begin{tabular}{p{5cm} p{2.0cm} p{2.0cm} p{2.0cm} p{2.0cm}}
\hline
Study & Accuracy & Macro-average F1 & Weighted-average F1 & Data Privacy \\
\hline
CA-ADP-3D-CNN-BiLSTM (proposed) & 0.970 & 0.89 & 0.97 & Yes \\
3D-CNN-BiLSTM (proposed, no DP) & 0.980 & 0.93 & 0.98 & No  \\
\cite{AL_QANESS_2024}            & 0.980 & 0.83 & 0.98 & No  \\
\hline
\end{tabular} 
\end{center}
\end{table}

\section{Discussion} \label{sec:discussion}

This study introduces a Class-Aware Adaptive Differential Privacy (CA-ADP) framework integrated with a hybrid 3D CNN-BiLSTM architecture. It is a privacy-preserving fall detection methodology using sensor data. Extensive experiments have been performed on three benchmark datasets: SisFall, UP-Fall, and MobiAct. We performed a statistical significance test, privacy–utility trade-off analysis, and a comparison of privacy guarantees and performance with previous studies. This article provides several important insights regarding the model architecture, class-aware privacy mechanisms, and classification performance.

The experimental results show that the hybrid 3D-CNN–BiLSTM (without DP) achieves the highest prediction performance across all evaluation metrics on the SisFall dataset. For the UP-Fall dataset, the 3D-CNN–BiLSTM outperforms the standalone 3D CNN and BiLSTM models in terms of all evaluation metrics. Similar trends were also observed on the MobiAct dataset. These findings indicate that the proposed 3D-CNN–BiLSTM model has the potential to achieve reliable prediction performance across the heterogeneous datasets.


When differential privacy was applied, we found a remarkable performance gap between the CA-ADP based 3D-CNN–BiLSTM model and the CONV-DP–based 3D-CNN–BiLSTM model. The experimental results show that the CA-ADP–based model consistently outperforms the CONV-DP–based model across all evaluation metrics and datasets. In conventional DP (DP-SGD), a fixed noise level is applied across all training samples, which may negatively affect the minority classes. In contrast, the proposed CA-ADP mechanism uses adaptive noise scaling strategy. It injects noise based on the class composition of the mini-batches. Batches with more fall samples (minority class) receive lower noise, while batches dominated by activities of daily living (ADL) receive higher noise. This adaptive approach helps preserve the important patterns associated with rare fall events, leading to better performance results.


The privacy–utility trade-off analysis presented in Section~\ref{sec:trade-off_analysis} shows that the relationship between noise and privacy is non-linear and varies from one dataset to another. In all three cases, CA-ADP achieves better privacy while improving utility. An optimal level of noise injection can effectively balance privacy and utility. In some cases,  CA-ADP even helps the model generalize better than the non-private version. For example, on the MobiAct dataset, CA-ADP achieved the highest recall (0.890), even surpassing the non-private hybrid model (0.851). This effect is especially important in fall detection, where datasets are heavily imbalanced. This adaptive mechanism prevents the model from overfitting to common daily activities (ADL) and improves its ability to detect less frequent fall events.


The Wilcoxon signed-rank test results confirm that the performance gains of CA-ADP over Conv-DP are statistically significant ($p < 0.0167$ after Bonferroni correction). When all 21 metric pairs from the three datasets are considered together, the pooled test gives $W = 1.5$ and $p = 0.0001$. This  provides strong evidence that the proposed mechanism consistently outperforms conventional differential privacy. The per-metric analysis shows that the most notable improvements occur in recall (+0.0883) and F1-score (+0.0643). These are especially important in fall detection, where missing a fall can have serious consequences. As a result, improvements in these metrics are not just statistically meaningful but also highly relevant for real-world applications.

To the best of our knowledge, this is the first study on a Class-Aware Adaptive Differential Privacy mechanism applied to the fall detection problem. Compared to existing fall detection studies, the proposed framework achieves competitive performance and provides formal privacy guarantees that none of the prior works have addressed. This analysis indicates that the proposed CA-ADP mechanism maintains near state-of-the-art results, demonstrating strong and reliable performance in a more realistic evaluation setting.

\section{Conclusions} \label{sec:conclusions}

This paper presented a privacy-preserving deep learning framework for sensor-based fall detection based on a hybrid 3D CNN--BiLSTM architecture integrated with a class-aware adaptive differential privacy mechanism. The proposed framework addresses a fundamental limitation of conventional DP-SGD, which applies fixed noise injection regardless of class distribution. Instead, the proposed mechanism dynamically adjusts the noise multiplier based on the class composition of each mini-batch. A formal $(\varepsilon, \delta)$-differential privacy guarantee was established and reported for all experiments. Experimental results confirmed that the CA-ADP mechanism consistently outperforms conventional differential privacy baselines in terms of accuracy, recall, F1-score, and ROC-AUC. The Wilcoxon signed-rank test with Bonferroni correction, discussed in Section~\ref{sec:statistical_test}, confirmed that these performance differences are statistically significant. The privacy-utility trade-off analysis further revealed that moderate noise injection can act as an implicit regulariser, in some cases improving generalisation relative to the non-private baseline. The comparison with prior studies showed that the proposed framework achieves competitive performance while providing formal privacy guarantees that none of the existing fall detection studies addressed. To the best of our knowledge, this is the first framework to apply class-aware adaptive differential privacy within a hybrid deep learning model for sensor-based fall detection, where class labels and mini-batches are used to adaptively tune noise injection for improved utility. 

Despite these promising results, several limitations should be acknowledged. The privacy budgets across all three datasets remain above $\varepsilon = 10$, falling short of the stringent guarantees advocated in the differential privacy literature. This is primarily a consequence of the small noise multiplier, the small batch size of 32, and the extended number of training epochs. In addition, all three datasets were collected under controlled laboratory conditions with scripted fall simulations, which may not faithfully represent the variability and unpredictability of real-world falls. Future work will pursue several directions to address these limitations. The privacy budget will be reduced toward the range of $\varepsilon = 3$ to $5$ through the use of larger noise multipliers and larger effective batch sizes. More advanced privacy accounting methods such as zero-concentrated differential privacy~\cite{bun2016concentrated} will also be explored. We also plan to extend the current framework beyond binary classification to support multi-class fall type recognition and severity estimation. Finally, integrating the CA-ADP mechanism within a federated learning framework offers a promising direction for distributed private training across wearable devices and clinical systems.

\section*{Declaration of Generative AI Use}
During the preparation of this work the author used ChatGPT and Claude in order to assist with language editing, grammar correction, and improving the readability of the manuscript. After using these tools/services, the author reviewed and edited the content as needed and take full responsibility for the content of the published article.

\bibliographystyle{elsarticle-num}

\bibliography{bibliography}

\end{document}